\documentclass[11pt]{article}
\usepackage{cmap}
\usepackage[T1]{fontenc}
\usepackage{lmodern}
\usepackage[margin=1in]{geometry}
\usepackage{amsmath,amssymb}
\usepackage{graphicx}
\usepackage{hyperref}
\usepackage{xcolor}
\usepackage{natbib}
\usepackage{booktabs}
\usepackage{siunitx}

\title{Does a Flux-Defined Markov Blanket Screen Self-Replicating Gray--Scott Spots?\\
\large A Computational Test of a Blanket Condition from the Free Energy Principle}
\author{John E. Pearson\\
Los Alamos, NM 87544}
\date{\today}

\newcommand{\eps}{\varepsilon}
\newcommand{\bB}{\boldsymbol{B}}


\newcommand{\movieurl}{https://www.youtube.com/shorts/qpzrgnDkdhY}
\begin{document}
\maketitle

\begin{abstract}
Friston's free energy principle holds that a system separated from its environment by a Markov blanket---a boundary whose states statistically screen its interior from its exterior---can be treated as performing approximate Bayesian inference, and that such blankets arise wherever coupling is short-range. We test this blanket condition in a concrete nonlinear pattern-forming system: self-replicating spots in a reaction-diffusion model that persist, compete, reproduce, and die. Directed boundary exchange and timescale separation hold up reasonably well. A flux-defined blanket does not screen interior from exterior. Flux removes only part of the linear dependence, and a nonlinear test shows that conditioning on flux \emph{increases} the dependence. We argue that this is because flux is a statistical \emph{collider}. Conditioning on the boundary's own concentration state screens about a fifth of the nonlinear dependence. The dependence is concentrated within about one organism radius of the boundary. A band of the surrounding field of that thickness does screen the interior from the exterior beyond it, as any short-range field would; this is not evidence of inference.
\end{abstract}


\section{Introduction}

\subsection{The Free Energy Principle and the Markov Blanket}
\label{sec:fep_intro}

Friston's free energy principle (FEP) argues that systems coupled by short-range forces should generically possess a \emph{Markov blanket}: a partition of the system's states into internal states $\mu$, external states $\eta$, and blanket states (sensory $s$ and active $a$), such that
\begin{equation}
p(\mu, \eta \mid s, a) = p(\mu \mid s, a)\, p(\eta \mid s, a), \label{eq:blanket_condition}
\end{equation}
i.e.\ internal and external states are conditionally independent given the blanket \citep{Friston2006,Friston2010,Friston2013,Friston2019Particular,Friston2023Simpler}. This structure is presented as generic: ``if the coupling among an ensemble of dynamical systems is mediated by short-range forces, then the states of remote systems must be conditionally independent'' \citep{Friston2013}.

The FEP's further claim is an inferential reading of the blanket: given a blanket, the internal states can be interpreted as performing approximate Bayesian inference about the external states, by minimising a variational free energy. The blanket condition is necessary but not sufficient for this reading. \citet{Friston2023Simpler} make the full logical structure explicit. The system, within the regime studied, must have a steady-state (time-invariant) density over its whole configuration, which, for our discretised two-field system, is a function of $2\times2^{20}$ variables; Eq.~\ref{eq:blanket_condition} must hold of that density; and a synchronisation condition must hold, so that the expected internal state given the blanket and the expected external state given the blanket are related by a smooth map. Only when all three hold does the inferential reading follow.

The framework has drawn substantial critical attention. \citet{Raja2021} argue that imposing a blanket partition after the fact licenses an inferential redescription of \emph{any} dynamics (the ``Markov blanket trick''), which makes the claims hard to falsify. \citet{Aguilera2022} show, for weakly coupled linear stochastic systems, that the blanket condition is not generic and requires sparse coupling structure. \citet{Bruineberg2022} argue that a purely statistical, Pearl-style Markov blanket \citep{Pearl1988} (a conditional-independence structure read off a joint distribution) is conflated in the FEP literature with Friston's dynamical blanket, which additionally constrains the flow of influence; we return to this in Section~\ref{sec:blanket_formulation}. Further critical literature is discussed in the Electronic Supplementary Material (ESM) \S\ref{sec:S_more_lit}.

Self-replication is of particular interest because reaction-diffusion spots are a fully specified system in which the question of cyclically reforming blankets can actually be posed (ESM \S\ref{sec:S_more_lit}). What is largely missing is a direct empirical test, in a nonlinear, spatially extended, many-body system, of whether an intuitively organism-like structure has the conditional independence the framework requires. Analytical constructions exist for low-dimensional systems such as two coupled Lorenz systems \citep{Friston2021Stochastic}. \citet{DaCosta2021} derive the inferential reading for stationary processes once a blanket partition is given, and \citet{Palacios2020} exhibit hierarchical self-organisation in simulations built on free energy minimisation. Organism-like patterns in cellular automata and other dynamical systems have been characterised through autopoiesis and information-based criteria \citep{Beer2015,Beer2020,Biehl2016}, which are not tests of blanket conditional independence. We are not aware of an earlier direct test of this kind in a spatially extended system of interacting, self-replicating structures, and this paper supplies one.

\subsection{The Gray--Scott Model and Self-Replicating Spots}

\begin{figure}[htbp]
\centering
% Clicking the image opens the movie (set \movieurl above). Border off for this link only.
\hypersetup{pdfborder={0 0 0}}%
\href{\movieurl}{\includegraphics[width=0.5\textwidth]{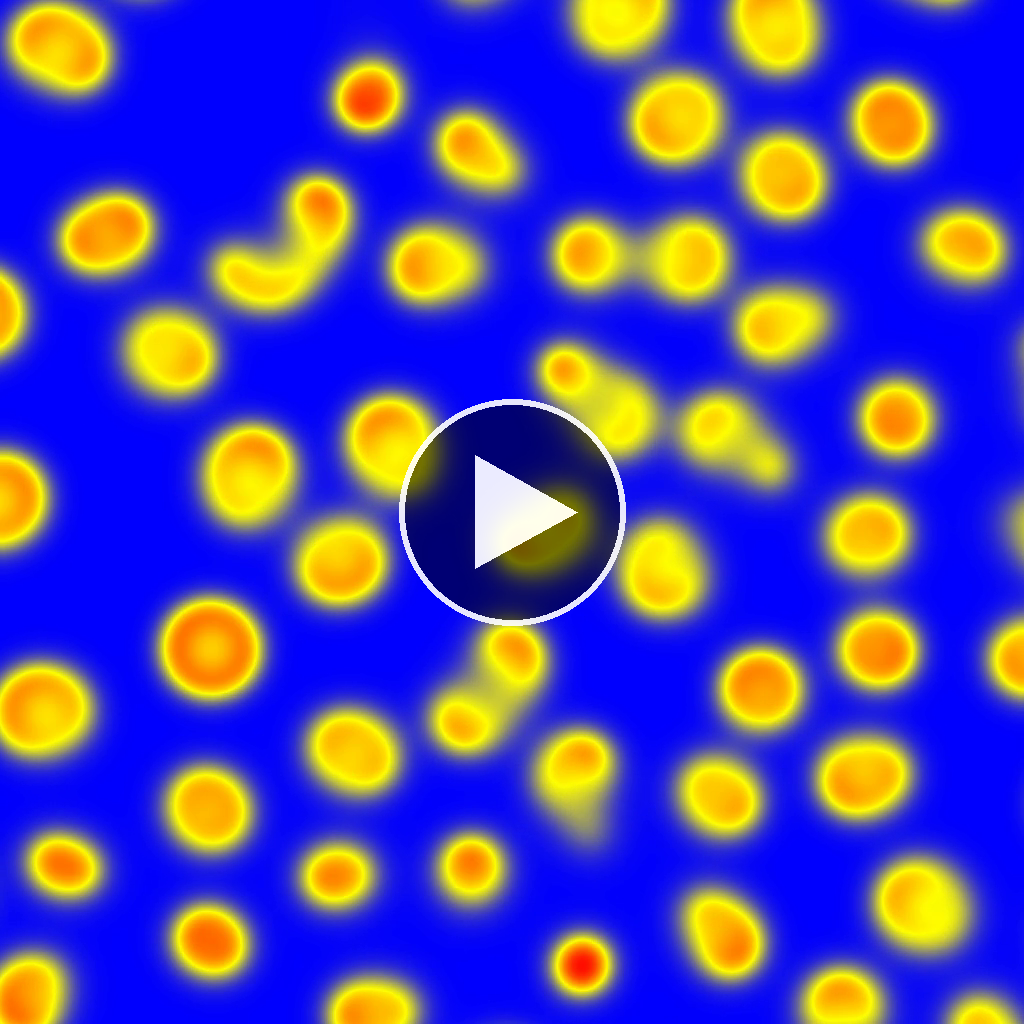}}%
\hypersetup{pdfborder={0 0 1}}
\caption{Activator field $v$ in the chaotic, spot-splitting regime studied here ($1024\times1024$): round spots, elongated organisms mid-division, and competitive packing. The image is a still frame from a movie of the simulation; click the image to play the movie in a web browser.}
\label{fig:spots_snapshot}
\end{figure}

The Gray--Scott system \citep{Gray1983,GrayScott1984} is a two-species reaction-diffusion model with the autocatalytic step $U + 2V \to 3V$. \citet{Pearson1993} showed that it supports isolated spots that grow, elongate and divide, reproducing through chemistry and diffusion alone. In crowded regimes the spots compete for substrate and die, producing sustained spatiotemporal chaos (Pearson's pattern class $\epsilon$; \citealt{Munafo}; Figure~\ref{fig:spots_snapshot}). The spots are compact, exchange material with their environment, and arise from a fully specified partial differential equation. If any simple system's spots should possess Markov blankets on Friston's own criteria, these seem good candidates.

Section~\ref{sec:model} defines the model and the individuation of organisms. Section~\ref{sec:blanket} tests three blanket conditions; the conditional-independence test of Section~\ref{sec:conditional_independence} is the primary result, and the later analyses test whether it survives. Section~\ref{sec:robustness} addresses the ergodicity assumption. Section~\ref{sec:freeenergy} constructs three notions of free energy as an exploratory control. Sections~\ref{sec:discussion} and~\ref{sec:conclusion} discuss and conclude.

\section{Model and Numerical Methods}
\label{sec:model}

We simulate the Gray--Scott system on a periodic square of side $L=2.5$:
\begin{align}
\frac{\partial u}{\partial t} &= D_u \nabla^2 u - u v^2 + F(1-u), \label{eq:gs_u}\\
\frac{\partial v}{\partial t} &= D_v \nabla^2 v + u v^2 - (F+k) v, \label{eq:gs_v}
\end{align}
with $D_u=2\times10^{-5}$, $D_v=10^{-5}$, and $F=0.018$, $k=0.055$ for the chaotic replicating regime \citep{Munafo}. Here $u$ (substrate) is fed at rate $F$ towards 1, $v$ (activator) is produced autocatalytically at rate $uv^2$, and $v$ decays at rate $F+k$. We use a $1024\times1024$ pseudo-spectral scheme (first-order exponential time differencing, $\Delta t=1$, one step per time unit), which resolves each organism at about 20 pixels per radius. Runs start from a pre-relaxed, populated field (ESM \S\ref{sec:S_reproducibility}). The scheme agrees closely with an independent $256\times256$ finite-difference scheme \citep{Pearson1993} and is insensitive to omitting an explicit de-aliasing rule (ESM \S\ref{sec:S_pseudospectral}).

\subsection{Individuation}
\label{sec:individuation}

Organisms are individuated geometrically from $v$ in four steps (ESM \S\ref{sec:S_individuation}, \S\ref{sec:S_merge_procedure}). (1) \emph{Threshold.} For every frame, threshold $v$ using Otsu's method \citep{Otsu1979}, making everything above the cutoff spot and everything below background. (2) \emph{Markers.} Take the local maxima of $v$ among the spot pixels, at least 24 pixels apart, as one marker per spot core, so that a spot with a slightly bumpy peak is not counted twice. (3) \emph{Watershed.} Turn $v$ upside down so that each spot is a bowl with its marker at the bottom, let every bowl fill at the same rate, and draw a boundary wherever water from two bowls meets, correctly treating spots that cross the periodic boundaries. This gives the \emph{raw regions}. (4) \emph{Merge.} Fuse touching raw regions whose equivalent-radius circles overlap, and repeat until none do, because the watershed can cut one physical spot into two pieces (ESM Figure~\ref{fig:merge_correction}). The merged result is the primary individuation used throughout.

\begin{figure}[htbp]
\centering
\includegraphics[width=\textwidth]{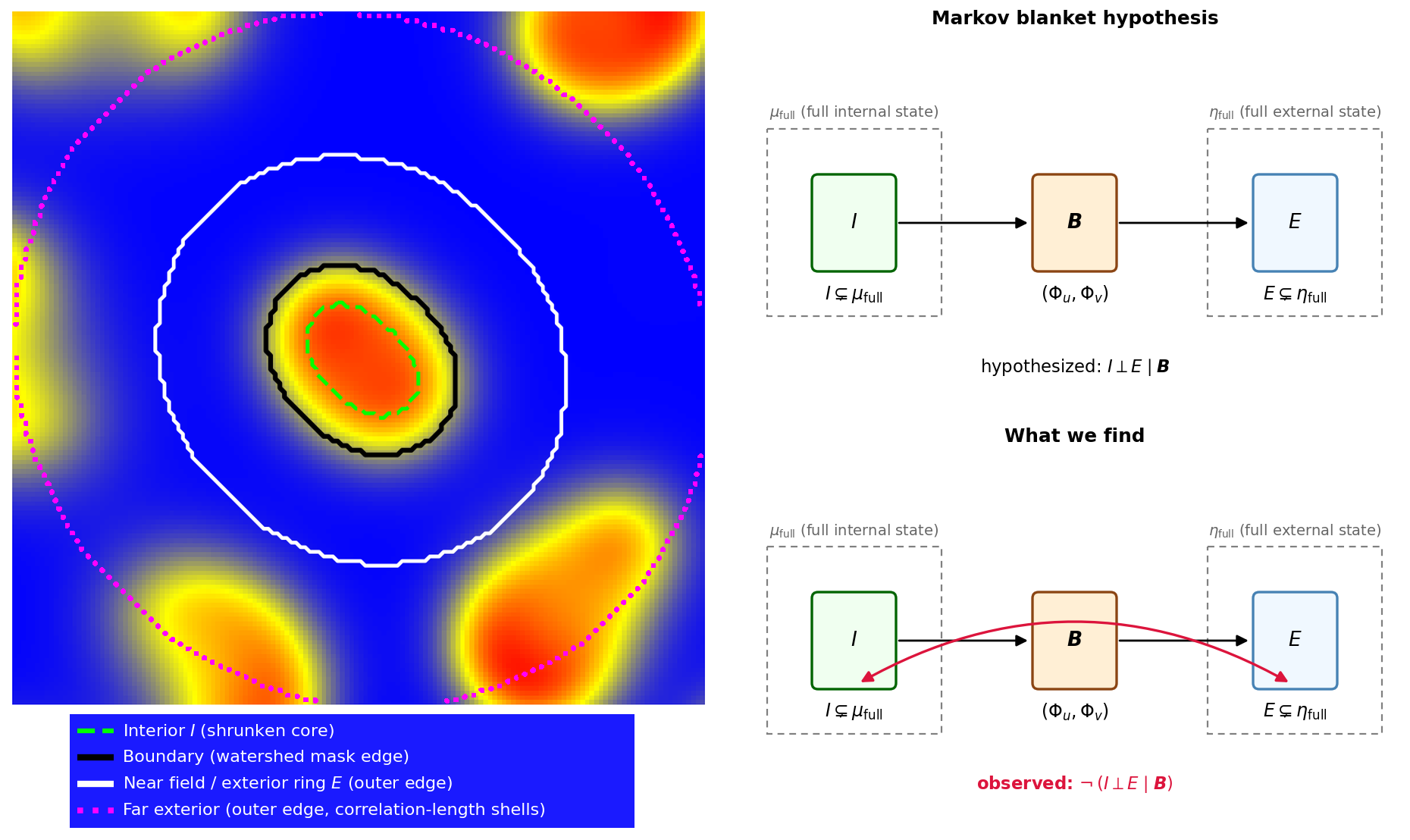}
\caption{Left: the test on a field snapshot: organism boundary (black), shrunken interior $I$ (green, dashed), outer edge of the exterior ring $E$ (white), and outer edge of the far exterior used in Section~\ref{sec:corrlength} (magenta, dotted). Right: the abstract structure. $I$ and $E$ are coarse-grained summaries of the full internal and external states, so a failure of screening between them does not by itself show that no partition of the full state space would screen. A blanket requires that conditioning on $\bB{}$ renders $I$ and $E$ independent (top); we find a persistent dependence (bottom).}
\label{fig:schematic}
\end{figure}

The \emph{interior state} $I$ is the mean $v$ in a core obtained by shrinking the organism's mask until each pixel is at least $50\%$ of the organism's radius from the outside (the result does not depend on this fraction; ESM \S\ref{sec:S_individuation}). The \emph{exterior state} $E$ is the mean $v$ in a ring $24$ pixels wide just outside the boundary (Figure~\ref{fig:schematic}, left). The organism's \emph{mask} is its segmented region and its \emph{boundary} is the contour of that mask. Section~\ref{sec:levelset} repeats the key test with a second, differently biased individuation method.

\section{Testing the Markov Blanket Hypothesis}
\label{sec:blanket}

Friston's criterion for a blanket requires three conditions, which we test in turn: \textbf{(A)} a separation of timescales, such that the blanket changes slowly relative to the states it couples \citep{Friston2013} (Section~\ref{sec:timescale}); \textbf{(B)} a directed sensory/active asymmetry (Section~\ref{sec:flux}); and \textbf{(C)} conditional independence of internal and external states given the blanket (Section~\ref{sec:conditional_independence}). Condition \textbf{(C)} is the load-bearing one; \textbf{(A)} and \textbf{(B)} are meant to make it possible but do not guarantee it. We take the blanket state to be the net diffusive flux $\bB{}=(\Phi_u,\Phi_v)$ of $u$ and $v$ across the boundary. \emph{Conditional coupling} means the dependence between $I$ and $E$ that remains after conditioning on $\bB{}$, measured by the partial correlation and by the conditional mutual information
\begin{equation}
\mathrm{MI}(X;Y\mid Z)=\iiint p(x,y,z)\,\ln\frac{p(x,y\mid z)}{p(x\mid z)\,p(y\mid z)}\,dx\,dy\,dz ,
\label{eq:cmi}
\end{equation}
in nats (the unit of information when logarithms are natural), which is zero if and only if $X$ and $Y$ are conditionally independent given $Z$ and, unlike a correlation, detects nonlinear dependence. $I$ and $E$ are coarse-grained, low-dimensional summaries of the full interior and exterior. The nearest-neighbour estimator used below could not be run on the full fields in any case, since nonparametric information estimation degrades sharply with dimension. This coarsening is licensed by a two-sided data processing inequality \citep{CoverThomas2006}: for functions $F,G$ applied separately to the two variables, $\mathrm{MI}(X;Y\mid Z)\ge \mathrm{MI}(F(X);G(Y)\mid Z)$. A positive $\mathrm{MI}(I;E\mid\bB{})$ is therefore a lower bound on the corresponding quantity for a richer description, so, for fixed masks, a positive finding cannot be an artefact of discarding information. The guarantee assumes the masks defining $I$ and $E$ do not depend on those values, a reasonable but unverified premise, and it gives no protection to the choice of $\bB{}$ itself, since conditioning is not processing in the inequality's sense (ESM \S\ref{sec:S_coarse}).

\subsection{Timescale Separation (Criterion A)}
\label{sec:timescale}

For organisms followed unambiguously between frames 10 time units apart, we compute $\rho = v_{\text{cntr}}\,\tau/R$, where $v_{\text{cntr}}$ is the centroid speed, $R$ the radius and $\tau = 1/(F+k) \approx 13.7$ the relaxation time of the featureless background state (ESM \S\ref{sec:S_jacobian}). Over $93{,}684$ transitions the median $\rho$ is $0.032$ (90th percentile $0.078$, 99th $0.16$): an organism moves about 3\% of its radius per relaxation time. Position is not shape, so we also compared each organism's boundary outline between matched frames after best-aligning rotation. Over $74{,}577$ matched transitions the median correlation is $0.916$. The lower tail largely reflects nearly circular organisms, whose shapes provide little basis for correlation. Measured directly, the boundary moves by a median of $3.0\%$ of the radius between frames (99th percentile $17\%$), and the largest changes occur where a boundary is made or lost, at divisions and before deaths (ESM \S\ref{sec:S_shape_tail}). Matched-asymptotic theory predicts slowly drifting spots only for small diffusivity ratios, far from ours ($D_v/D_u=0.5$); we do not claim equivalence with it, only that the separation persists (ESM \S\ref{sec:S_more_lit}).

Friston's criterion compares the blanket's rate of change with that of the states it separates. The lag at which each autocorrelation first falls below $1/e$ is about $48$ time units for $I$, $62$ for $E$, and $61$ and $58$ for $\Phi_u$ and $\Phi_v$, all well above $\tau$. The blanket is slower than $I$ but not slower than $E$, so the strict form of the criterion remains unestablished in either direction (ESM \S\ref{sec:S_autocorr}).

\subsection{Exchange-Flux Asymmetry (Criterion B)}
\label{sec:flux}

By the divergence theorem the net outward diffusive flux of species $X$ is $\Phi_X = -D_X \int_\Omega \nabla^2 X\, dA$, computed from the discrete Laplacian. Across about $96{,}000$ organism-instances, $\Phi_u<0$ (net inward) and $\Phi_v>0$ (net outward) in all of them: the substrate $u$, which neighbouring organisms compete for, flows in, and the activator $v$ flows out. This is not strong independent evidence of a directed sensory/active structure, since an organism is individuated as a region of high $v$ and depleted $u$, so diffusion down each gradient is nearly a consequence of the segmentation. Diffusive flux also depends on concentrations on both sides of the boundary (Fick's law), so each channel is coupled in both directions, a weaker kind of ``action'' than critics argue can license an inferential interpretation \citep{Raja2021}. The check shows that the labelling is coherent, not that the dependency structure the framework needs is present.

\subsection{Conditional Independence (Criterion C)}
\label{sec:conditional_independence}

We test whether $I \perp E \mid (\Phi_u, \Phi_v)$. This concerns the operational partition above and is not a search over arbitrary partitions of the full state space. The unit of analysis is the organism-instance, one organism in one sampled frame. Instances in the same frame share a field and are not independent, so we also report intervals from a bootstrap that resamples whole frames, and we rely on effect sizes as well as significance.

\subsubsection{Linear partial correlation}
\label{sec:ci_ensemble}
\label{sec:ci_single_traj}
\label{sec:ci_single_traj_run}

We regress $I$ and $E$ separately on $\bB{}_i=(1,\Phi^{\text{PC1}}_i)$ by least squares, where $\Phi^{\text{PC1}}$ is the leading principal component of the standardised flux pair, and correlate the residuals $\Delta I_i$, $\Delta E_i$:
\begin{equation}
r = \frac{\sum_{i} \Delta I_i\, \Delta E_i}{\sqrt{\sum_{i} \Delta I_i^2 \sum_{i} \Delta E_i^2}}.
\label{eq:partial_corr}
\end{equation}
We use PC1 because $\Phi_u$ and $\Phi_v$ are nearly collinear ($r\approx-0.98$). Regression on both fluxes together (the joint regression) is numerically well determined here but gives a larger value, outside the range of the two single-flux values; we report PC1 as the smaller, more conservative estimate and treat the joint value as a further sign of the collider structure discussed below (ESM \S\ref{sec:S_pc1}).

To avoid treating snapshots of one run as independent, we also apply the test to three independent ensembles of realisations (Table~\ref{tab:ensemble_summary}). Each realisation is branched from one converged configuration, perturbed once by Gaussian noise ($\sigma=10^{-3}$, distinct seed), and evolved for a burn-in with no further noise. By the end of a $3000$-step burn-in the fields are close to decorrelated (their difference is $68$--$84\%$ of that between independent draws of the field, and $98$--$107\%$ at $5000$ steps; ESM Figure~\ref{fig:lyapunov}), but the organism count takes longer (Section~\ref{sec:robustness}), so the third ensemble uses a $5000$-step burn-in and wider sample spacing ($1000$ steps rather than $100$; ESM \S\ref{sec:S_ensemble2}). The table also gives the single trajectory ($N=96{,}315$ instances from $500$ snapshots of one long run), which alone records the shell profiles and lineages the later analyses need.

\begin{table}[htbp]
\centering

\footnotesize
\setlength{\tabcolsep}{3.5pt}
\begin{tabular}{lrrrccccc}
\toprule
Dataset & Runs & Burn-in & $N$ & Marginal $r$ & Partial $r$ & Screened & Naive CI & Resampled CI \\
\midrule
Ensemble 1 & $1000$ & $3000$ & 566{,}689 & $-0.233$ & $-0.1815$ & $39.1\%$ & $[-0.1840,-0.1790]$ & $[-0.1850,-0.1781]$ \\
Ensemble 2 & $1000$ & $3000$ & 566{,}012 & $-0.235$ & $-0.1843$ & $38.3\%$ & $[-0.1868,-0.1818]$ & $[-0.1877,-0.1809]$ \\
Ensemble 3 & $500$ & $5000$ & 288{,}286 & $-0.244$ & $-0.1889$ & $40.2\%$ & $[-0.1924,-0.1854]$ & $[-0.1941,-0.1839]$ \\
Trajectory & --- & --- & 96{,}315 & $-0.240$ & $-0.1896$ & $37.7\%$ & $[-0.1957,-0.1835]$ & $[-0.1992,-0.1803]$ \\
\bottomrule
\end{tabular}
\caption{Interior--exterior correlation before and after conditioning on flux (PC1), with 95\% confidence intervals. Runs are realisations, and burn-in is in time steps after branching. The trajectory is the single long run, sampled every $100$ steps from $t=3900$, after the $3800$ steps of growth of the settled field all runs start from. ``Screened'' is $1-r_{\mathrm{partial}}^2/r_{\mathrm{marginal}}^2$. The naive interval treats organisms as independent. The resampled interval does not: it resamples whole frames (2000 replicates) for the trajectory and whole realisations for the ensembles.}
\label{tab:ensemble_summary}
\end{table}

The estimates agree in sign and magnitude, which we take as weak, informal support for the ergodicity assumption. Of the $1000$ realisations in the first ensemble, $999$ have the pooled estimate's sign. At these sample sizes rejecting exact conditional independence is nearly guaranteed, so the substantive quantity is the screened fraction: flux removes about $38$--$40\%$ of the marginal shared variance, real but well short of the near-complete screening a blanket requires.

\subsubsection{Nonlinear conditional mutual information}
\label{sec:ci_cmi}

A linear test cannot detect purely nonlinear dependence, so we also estimate $\mathrm{MI}(I;E\mid\bB{})$ with the Frenzel--Pompe nearest-neighbour estimator ($k_{\mathrm{NN}}=5$; \citealt{Kraskov2004,FrenzelPompe2007}), checked on synthetic data with known ground truth and against an independent implementation (ESM \S\ref{sec:S_cmi_calibration}). $I$, $E$, $\Phi_u$ and $\Phi_v$ are each $z$-scored (rescaled to zero mean and unit variance), applied identically to the observed statistic and every null trial, because the estimator is scale sensitive when inputs span very different ranges, as flux ($\sim10^{-4}$) and $I$, $E$ ($\sim10^{-1}$) do (ESM \S\ref{sec:S_cmi_calibration}). This estimator conditions on both fluxes jointly; it involves no regression, so their collinearity causes no instability.

\begin{figure}[htbp]
\centering
\includegraphics[width=0.85\textwidth]{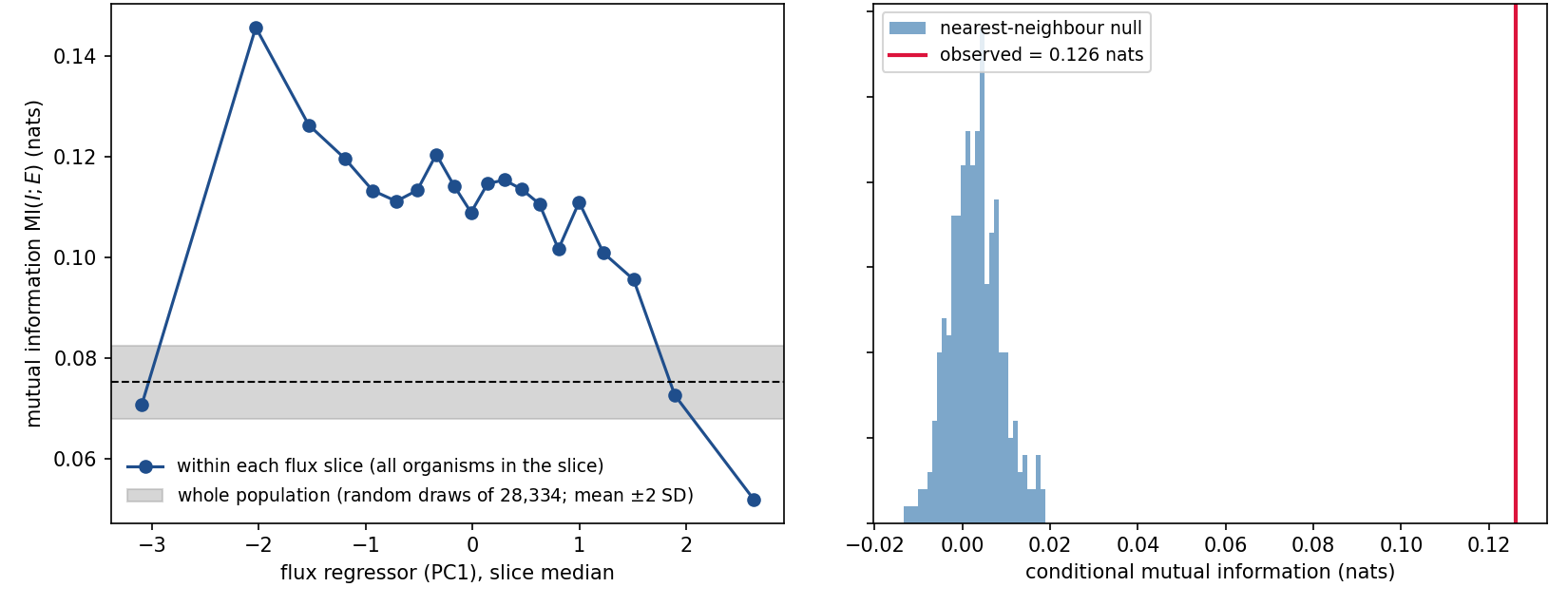}
\caption{Nonlinear interior--exterior dependence in the independent ensemble ($N=566{,}689$). Left: organisms are sorted by flux (PC1) and divided into 20 groups of equal size, so that organisms within a group have similar flux. Each dot is $\mathrm{MI}(I;E)$ estimated on all of the roughly $28{,}000$ organisms in a group, plotted at the group's median flux. The grey band is the same estimate from random draws of $28{,}334$ organisms from the whole population (mean $\pm2$ SD of a single estimate over 20 draws; the mean, dashed in the figure, is $0.075$ nats). Flux would screen if every group sat near zero; instead 17 of the 20 groups lie above the band, which is itself well above zero. Right: observed CMI on one $N=8000$ subsample (red line) against the nearest-neighbour conditional null (blue), $21.5$ null SDs above its mean.}
\label{fig:ensemble_summary}
\end{figure}

On $N=8000$ subsamples of the single trajectory the CMI given flux is $0.130$ nats, against $0.075$ nats unconditioned, so conditioning on flux \emph{raises} the dependence by about $73\%$ (Table~\ref{tab:cmi}). The estimate varies from subsample to subsample, so the table gives means over $20$ subsamples. Each is tested against a null built by local permutation among the organisms with the nearest fluxes in the style of \citet{Runge2018} ($300$ trials); in individual subsamples the conditioned values lie between $21$ and $32$ null standard deviations above it. The independent ensemble agrees: three subsamples of $8000$ give $0.147$ nats given flux ($22$--$27$ null SDs; Figure~\ref{fig:ensemble_summary}, right, shows one of them). We checked both this null and a second one, which shuffles $E$ only among organisms in the same one of $30$ groups of similar flux (found by $k$-means clustering), on $48$ synthetic replicates in which $I\perp E\mid\bB{}$ holds by construction. The null-SD statistic had mean $0.03$ and SD $0.92$ for the local null ($0.07$ and $0.97$ for the grouped null), and about $6\%$ of replicates exceeded $1.645$ for either, close to the nominal $5\%$ (ESM \S\ref{sec:S_cmi_calibration}). We still read the figures as diagnostic comparisons, since the check uses a constructed independence. Inside slices of fixed flux the dependence exceeds that of the whole population in $17$ of $20$ slices, peaking at $1.94$ times the unconditioned value in the second-lowest-flux slice, and falls to the unconditioned value or below only in the lowest-flux slice and the two highest (Figure~\ref{fig:ensemble_summary}, left).

\begin{table}[!t]
\centering
\small
\begin{tabular}{lcc}
\toprule
Condition on & CMI (nats) & Dependence \\
\midrule
nothing (marginal $\mathrm{MI}(I;E)$) & $0.075\pm0.002$ & --- \\
flux $(\Phi_u,\Phi_v)$ & $0.130\pm0.001$ & up $73\%$ \\
boundary concentration $(u_\partial,v_\partial)$ & $0.061\pm0.002$ & down $19\%$ \\
\bottomrule
\end{tabular}
\caption{Nonlinear interior--exterior dependence in the single trajectory ($N=96{,}315$): means $\pm$ standard error over $20$ subsamples of $8000$. The last column is the change from the marginal dependence. Conditioning on flux raised the dependence in all $20$ subsamples.}
\label{tab:cmi}
\end{table}

Flux is computed from the concentrations on both sides of an organism's own boundary: from the interior side, which $I$ summarises, and from the exterior side, which $E$ summarises. A diffusive flux depends on the difference between the two. The structure is therefore $I\rightarrow\Phi\leftarrow E$: flux is a joint downstream effect of the two sides, not a state either side acts \emph{through}. That is a statistical collider, and conditioning on a collider couples its two sources instead of separating them. Once the flux is fixed, a change in one side's concentration has to be accompanied by a change in the other's, so knowing $I$ again tells us something about $E$, even where the two would otherwise be less related (Section~\ref{sec:blanket_formulation}). We tested this by conditioning instead on the organism's own boundary concentration $(u_\partial,v_\partial)$, a genuine state. Flux raises the dependence, as a collider should; boundary concentration lowers it, by about a fifth (Table~\ref{tab:cmi}). The two move in opposite directions and neither comes near the screening a strict blanket requires. We keep flux as the blanket variable throughout, since swapping variables for the best-looking result would be the post-hoc partition fitting \citet{Raja2021} warn against (ESM \S\ref{sec:S_collider_test}).

\subsubsection{Alternative explanations}
\label{sec:alt_explanations}

We checked five less interesting explanations: local crowding, a diffuse whole-domain confound, life-cycle stage, a between-organism mixture artefact, and organism size and shape with frame-wide fluctuations. None explains the violation away (ESM \S\ref{sec:S_alt_explanations}). Local crowding, measured as the number of neighbours within 80 pixels, leaves the correlation of the same sign and slightly stronger ($r=-0.226$ against $-0.190$), and subtracting each frame's average from every variable, which removes any frame-wide influence, changes it by about $4\%$. A lineage is one organism followed from frame to frame, from its birth by division to its own death or division. Comparing each organism only with its own lineage, by subtracting the lineage's average from every variable, still gives $r=-0.261$, with all of the covariance within lineages, ruling out a pure mixture of organisms that differ from one another. The coupling is weakest in the youngest third of organisms ($r=-0.097$) and about four times stronger in the older thirds, so it builds with age rather than being inherited. Organism size alone does not account for it (marginal $r$ $-0.240\to-0.234$). Adding radius and eccentricity \emph{to the flux conditioning} flips the sign ($r=+0.248$), but we read that as a limit of conditioning on flux: radius shares about $70\%$ of its variance with the first principal component of flux, so the adjusted flux is nearly a readout of $I$ itself. Among the $89{,}689$ organisms ($93.1\%$) that individuate without merging, $r=-0.184$ against $-0.190$ overall.

\subsubsection{Robustness to a non-watershed individuation method}
\label{sec:levelset}

Watershed boundaries are drawn relative to neighbours, so the violation might be an artefact. We repeated the test with fixed-threshold connected components of $v$, which share no construction with watershed, on $100$ archived frames. Excluding touching pairs gives $r=-0.259$ ($N=15{,}970$); recovering them with a local watershed split gives $r=-0.068$ ($N=19{,}593$). The sign and a clearly nonzero value survive, but the linear magnitude is about a third of the watershed value. The nonlinear test agrees more closely: conditioning on flux raises the dependence from $0.086$ to $0.148$ nats on the pooled sample ($24$ null SDs), as for watershed organisms. The recovered pairs alone are the exception, where flux removes most of a small dependence ($0.039\to0.017$ nats, $2.1$ SDs; ESM \S\ref{sec:S_levelset}).

We conclude that the conditional-independence property of a Markov blanket, as formalised by \citet{Friston2013}, does not hold for the flux-defined partition of individual organisms in this system, across linear and nonlinear estimators, single-trajectory and ensemble sampling, and two individuation methods.

\subsection{Spatial Structure of the Conditional Coupling}
\label{sec:corrlength}

\begin{figure}[htbp]
\centering
\includegraphics[width=0.75\textwidth]{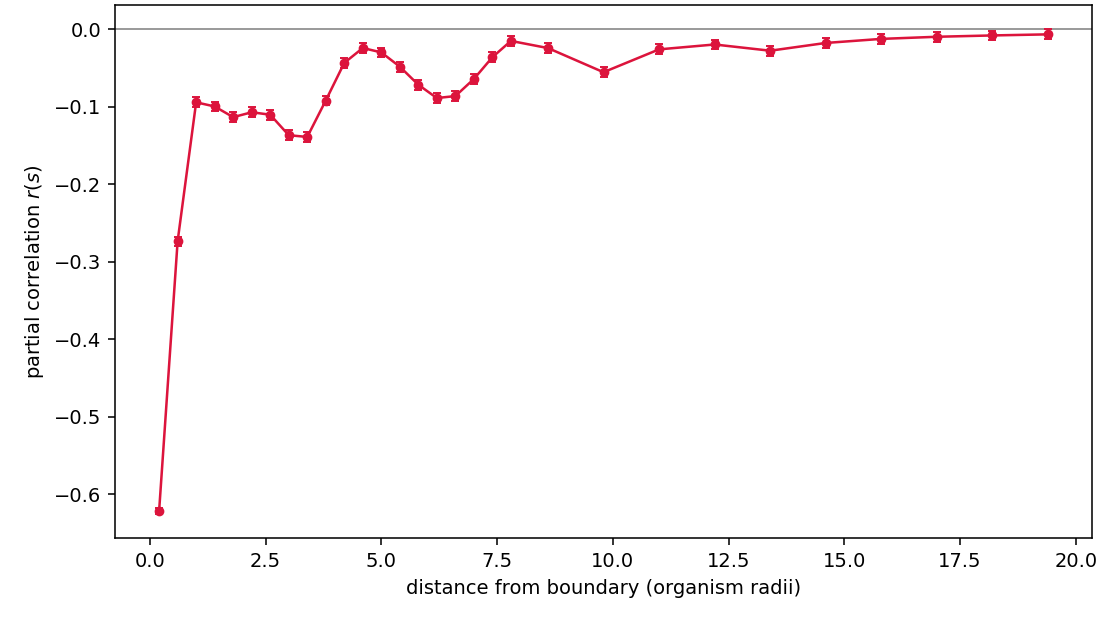}
\caption{Partial correlation (PC1 flux regressor) between interior and exterior state as a function of distance from the organism's boundary in organism radii ($N=96{,}315$; error bars 95\% CI).}
\label{fig:decay_curve}
\end{figure}

We measured $r(s)$, Eq.~\ref{eq:partial_corr} with $E_i$ replaced by the mean $v$ in the organism's own concentric shell at distance $s$ from its boundary (shells $8$ px wide near the organism, $24$ px wide farther out), out to 19--20 organism radii ($N=96{,}315$; ESM \S\ref{sec:S_corrlength}; PC1 regressor, since the joint estimate falls outside the single-flux range at 29 of 30 shells, ESM \S\ref{sec:S_pc1}). The curve (Figure~\ref{fig:decay_curve}) is negative at every distance tested. The coupling is concentrated within about one radius of the boundary but does not vanish at the greatest distances.

\subsection{Richer Blanket Representations}
\label{sec:richer_blanket}

A linear enrichment, flux plus the organism's own $v$ in $16$ concentric exterior shells about $8$ px wide, does not screen the far-field exterior; adding the near-field shells reverses the sign of the correlation, which arises from near-collinear predictors, not from screening (ESM \S\ref{sec:S_richer_blanket}). We also tested the shared substrate $u$ nonlinearly. Conditioning on flux plus the mean $u$ in seven shells outside the boundary lowers the CMI from $0.134$ to $0.001$ nats, much more than seven columns of random noise do ($0.046$). But this is not screening. The shells lie inside the ring over which $E$ is measured, so they largely contain $E$: they explain $86$--$92\%$ of its variance, and conditioning on a noisy copy of $E$ that explains the same fraction lowers the CMI nearly as much (ESM \S\ref{sec:S_u_enriched}).

Because those shells overlap $E$, we asked how much of the surrounding field must be conditioned on before the interior is independent of the exterior beyond it. We took a band of width $w$ beyond the boundary, summarised by three area-weighted means each of $v$ and $u$, and measured the dependence between the interior and the $24$-px shell \emph{beyond} the band, so that nothing in the conditioning set overlaps the exterior variable (specification fixed in advance; ESM \S\ref{sec:S_band}). With flux alone ($w=0$) the dependence is $0.173$ nats; this exceeds the headline $0.130$ because $E$ here is computed from the concentric-shell profiles, not from the $24$-px ring of Section~\ref{sec:individuation}. The dependence falls as the band thickens, and from $w=24$ px, about one organism radius ($R\approx21$ px), outward it is at most $0.005$ nats, under $5\%$ of the flux-only value and at most $3.1$ null SDs (Table~\ref{tab:band}). A noisy copy of the exterior variable screens as well, as it should in a spatially continuous field. A band about one radius thick therefore screens the interior from the exterior beyond it, and the thin boundary-defined blanket still fails.

Screening is not specific to organisms. As a control, we repeated the sweep on disks of the same size placed where no organism is. They screen too, and sooner: from $16$ px they are no more than $0.003$ nats above the null, whereas the organisms' dependence persists further out (Table~\ref{tab:band}).

\begin{table}[htbp]
\centering
\small
\begin{tabular}{rcc}
\toprule
Band width $w$ (px) & Organisms & Placebo disks \\
\midrule
$0$ (flux only) & $0.173$ & $0.217$ \\
$8$ & $0.092$ & $0.018$ \\
$16$ & $0.032$ & $-0.008$ \\
$24$ & $0.000$ & $-0.006$ \\
$40$ & $0.004$ & $-0.006$ \\
$64$ & $0.001$ & $-0.002$ \\
$96$ & $-0.002$ & $-0.002$ \\
\bottomrule
\end{tabular}
\caption{Dependence between the interior and the $24$-px shell beyond a band of width $w$, given flux and summaries of the band (CMI, nats). Organisms: mean of three subsamples of $8000$. Placebo disks: disks of matched size placed where no organism is, one subsample of $8000$ (ESM \S\ref{sec:S_band}). Values at or below zero are estimator noise.}
\label{tab:band}
\end{table}

A chemical memory variable added to the model does not remove the violation (partial correlation in the innermost shell $-0.58$ against $-0.62$), and when it does not diffuse the pattern freezes into a lattice (ESM \S\ref{sec:threevar}).

\section{Robustness of the Statistical Approach}
\label{sec:robustness}

The single-trajectory analyses treat time-separated samples of one chaotic trajectory as approximately independent draws from a stationary distribution, an ergodicity assumption we cannot prove. Three claims should be kept apart: that the system stays in a statistically stable regime, that our statistics are robust to initialisation and sampling, and that the sampling estimates the invariant distribution the theory refers to. Our checks bear on the first two (ESM \S\ref{sec:S_robustness}). Perturbed copies diverge until their fields differ as much as independent draws would. Measured from three reference states of one long run ($10{,}000$, $319{,}000$ and $469{,}000$ steps after the settled starting field), the time to reach that level agrees to within about $10\%$ ($3300$--$3600$ steps), and the growth rates of the divergence agree to within about $30\%$ ($0.0024$--$0.0033$ per time unit). Small-amplitude noise everywhere, with no localised seed, relaxes to the trivial state $(u,v)=(1,0)$ rather than to the patterned regime, so the patterned regime has a distinct basin that our seeded runs never leave (ESM \S\ref{sec:S_robustness}). One check qualifies this: in $12$ noisy copies run for $15{,}000$ steps, the organism count is still correlated across copies at the $3000$-step burn-in ($r=0.46\pm0.16$), though the correlation is indistinguishable from zero by about step $4800$. The copies are therefore not fully independent early on, and confidence intervals that pool them are somewhat too narrow; the spread across realisations (ESM \S\ref{sec:S_ensemble2}) is the better guide to uncertainty. The estimate itself is not affected: a third ensemble with a $5000$-step burn-in gives the same partial correlation ($r=-0.189$, against $-0.182$ and $-0.184$ for the first two). The headline results are computed per organism, not from the population count. Instances from the same frame are also not independent. The whole-frame resampling intervals in Table~\ref{tab:ensemble_summary} are wider but still decisively nonzero, and resampling blocks of up to $40$ consecutive frames changes them little. The nonlinear CMI is also unchanged on subsamples with one organism per frame or organisms at least ten radii apart (ESM \S\ref{sec:S_robustness}).

\section{Free Energy}
\label{sec:freeenergy}

Having found that the natural partition does not satisfy the blanket condition, we ask whether a well-defined free energy can still be constructed for these organisms and whether it recovers the life-cycle behaviour that motivated the work: rising towards division and falling after. This section is an exploratory control, not a further test of the blanket hypothesis: all three quantities are computed from the same fields that define the organisms. It guards against the inference that a well-behaved variational free energy would itself demonstrate inference.

\subsection{Template-Deviation Free Energy}
\label{sec:template_fe}

The simplest, purely descriptive proxy compares an organism's radial $v$-profile $v_i$ with an empirical template $T(r)$, the mean profile over $96{,}315$ organisms of the single trajectory ($20$ bins over $0$--$80$ pixels; ESM \S\ref{sec:S_template_procedure}):
\begin{equation}
\mathcal{F}_{\text{template}} = \frac{1}{n_{\text{bins}}}\sum_{i} \left( v_i - T(r_i) \right)^2 .
\end{equation}
It needs no generative model, prior or posterior, only a distance from a typical shape, chosen for direct comparability with the quantity below. The profile has a sensible form: near-maximal at the centre ($v\approx0.28$), falling to a minimum at about the organism radius plus the gap to the background, and rising as neighbours' peaks enter the average. An organism far from the template is typically small and over-concentrated at its centre, plausibly a recently divided daughter not yet at full size.

\subsection{Friston Free Energy}
\label{sec:friston_fe}

Variational free energy is a functional of an assumed generative model and an approximate posterior, not a statement about a system's causal structure:
\begin{equation}
\mathcal{F} = \mathrm{KL}\big(q(s) \,\|\, p(s)\big) - \mathbb{E}_{q(s)}\big[\log p(o \mid s)\big]. \label{eq:fe_definition}
\end{equation}
Here $s$ is the hidden state, $o$ the observation, $p(s)$ the prior over $s$, $q(s)$ the approximate posterior, and $\mathrm{KL}$ the Kullback--Leibler divergence, a measure of how far one distribution is from another. We instantiate it from the data of Section~\ref{sec:conditional_independence} with linear-Gaussian fits: hidden state $s:=E$; observation $o:=(\Phi_u,\Phi_v)$; prior $p(s)$ fit to the marginal of $E$; likelihood $p(o\mid s)=\mathcal{N}(As+b,\Sigma)$ fit by regression; and approximate posterior $q(s\mid I)$ a Gaussian whose mean is a regression of $E$ on the organism's own $I$ (ESM \S\ref{sec:S_surprisal}). Calling $q(s\mid I)$ an ``approximate posterior'' invites three readings, and only the first is intended. (i) It is a mathematical ingredient, a distribution chosen so that $\mathcal{F}$ is well defined and computable. (ii) It is not a posterior given $o$ in the Bayesian sense. Following Friston's synchronisation map, the claim that internal states parameterise beliefs about external states, we make it a function of the internal state $I$, not of the observation $o$. (iii) We make no claim that the dynamics compute it: Eqs.~\ref{eq:gs_u}--\ref{eq:gs_v} contain no reference to $I$, $E$ or $\Phi$. Since $\mathcal{F}$ can be computed identically whether or not the organism performs inference, a well-behaved $\mathcal{F}$ cannot by itself support that reading.

Fit on the ensemble, the KL term is small but positive (mean $0.028$), reflecting marginal dependence, not the conditional-independence violation. Refitting on a random half of the data and evaluating on the held-out half gives $r=1.000000$. Because all distributions are Gaussian, $\mathcal{F}=-\ln p(o)+\mathrm{KL}\big(q(s\mid I)\,\|\,p(s\mid o)\big)$ exactly, which we confirmed on $52{,}848$ life-cycle instances. The KL remainder is small (median $0.010$ nats, 95th percentile $0.09$, against $\mathcal{F}\approx-22$ in the arbitrary flux units), so $\mathcal{F}$ is, to a very good approximation, the surprisal of the organism's boundary flux under the population's flux statistics; through the life cycle the two curves correlate at $r=1.000$ and the remainder rises from about $0.03$ nats long before division to about $0.10$ shortly before it (ESM \S\ref{sec:S_surprisal}). This is a statement about the statistics of the flux, not about the organism being surprised.

\subsection{Chemical Free Energy and the Limits of Its Definition}
\label{sec:chemical_fe}

We also asked whether a genuine \emph{thermodynamic} free energy or entropy production rate can be defined. It cannot without a further modelling choice, and we do not attempt to supply one. The autocatalytic step is irreversible, so with a small reverse rate $k_r$ the affinity diverges as $-\ln k_r$ and no well-posed entropy production exists at $k_r=0$ (\citealt{Mahara2004}, who reversibilise both reactions, make this choice explicitly). We therefore do not compute an absolute value and follow instead the shape of the reaction-rate prefactor $uv^2$, to which $\sigma/(-\ln k_r)$ converges as $k_r\to0$, as a small-$k_r$ leading-order comparison and not an identity.

\subsection{Comparison of Life-Cycle Trajectories}
\label{sec:fe_comparison}

\begin{figure}[htbp]
\centering
\includegraphics[width=0.8\textwidth]{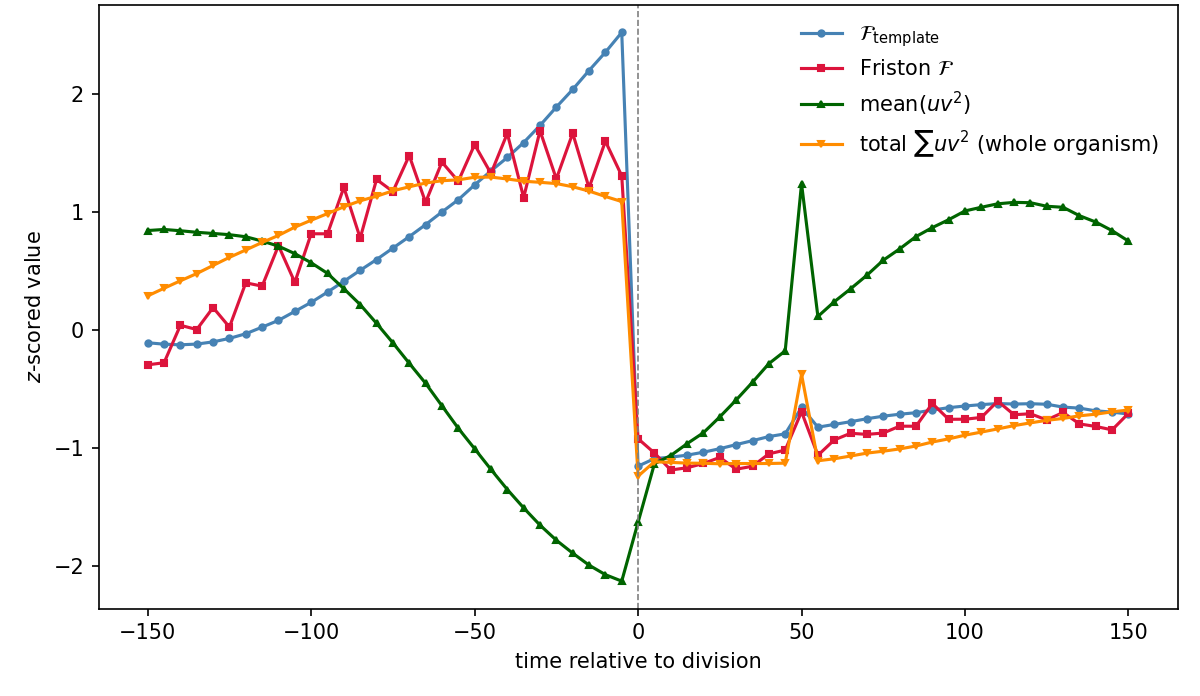}
\caption{Free energies and reaction rate aligned to division and averaged over $N=689$ confirmed divisions, each $z$-scored. Template-deviation and Friston free energy track closely ($r=0.934$). The per-area mean of $uv^2$ runs opposite; the total of $uv^2$ follows them but is set mainly by size.}
\label{fig:fe_comparison}
\end{figure}

Organisms are followed with a lineage tracker. Watershed segmentation is purely geometric and cannot distinguish a division from a spot that transiently deforms into a peanut shape and relaxes, so a one-becomes-two event counts as a division only if the children persist beyond a $50$-time-unit window (ESM \S\ref{sec:S_lineage}). The tracker follows the same fused organisms as the cross-sectional analyses. Median lifespans are $555$ (death) and $1265$ (division) time units, longer than the $\sim660$ reported by \citet{Pearson1993} for an isolated spot, as expected when neighbours compete for substrate. Over $N=689$ confirmed divisions, aligned to each division and averaged, $\mathcal{F}_{\text{template}}$ rises from $\approx0.0014$ to $\approx0.0031$ just before division, falls more than threefold at division, and rises again. The Friston $\mathcal{F}$ has a similar shape (Figure~\ref{fig:fe_comparison}). The reaction-rate prefactor $uv^2$ gives a third curve in two forms. As a per-area mean it runs \emph{opposite} to the free energies, falling to a minimum just before division and jumping at division, because a growing organism adds mostly low-activity interior to its mask. As a total over the organism it follows them closely but tracks size. Its close agreement with the Friston $\mathcal{F}$ is expected: $\mathcal{F}$ is built from the boundary flux of $u$, and by conservation of $u$ the inward flux equals the reaction's consumption in the organism less the feed supplied there; the two correlate $r=0.99$ over $326{,}315$ organism-instances (ESM \S\ref{sec:S_uv2_flux}).

Two checks address artefacts of the agreement (ESM \S\ref{sec:S_fe_comparison}). Rebuilding the template from a separate trajectory preserves it, and the agreement holds at the organism-instance level with no averaging ($r=0.863$, $N=95{,}805$). Both quantities track organism size (each about $-0.23$ with equivalent radius), nonlinearly. Controlling for age leaves the correlation unchanged and a linear size control barely changes it, but a cubic function of equivalent radius lowers it to $0.71$. About a third of the shared variance is therefore size; the rest is a strong agreement between two functions of the same measured state, which does not by itself show that either is inference.

The curves resume the same rising trend in the daughter generation rather than restarting from an unrelated baseline, consistent with something of the parent's state carrying over; we have not tested whether a parent's value at division predicts its daughter's starting value. Both quantities show the life-cycle pattern the original hypothesis predicted, and they agree despite very different derivations. That is a robust, quantitative life-cycle signature. It is not evidence that these organisms perform approximate Bayesian inference. The dynamics are fully determined by Eqs.~\ref{eq:gs_u}--\ref{eq:gs_v}, and finding a generative model under which $\mathcal{F}$ behaves as it would in an inferential system is not evidence of inference: a fit counts only if it could have failed were the hypothesis false, and a flexible model fit to almost any smooth, roughly Gaussian system will find some consistent parameterisation. What would count is the property the fit does not test, that internal and external states are screened through the blanket and nothing else. Section~\ref{sec:conditional_independence} tested that and found it false for the boundary-defined blankets; a thick band of the field screens, but that is a property of the field (Section~\ref{sec:richer_blanket}).

\section{Discussion}
\label{sec:discussion}

\subsection{Which Parts of the Framework Hold}

The three conditions fare differently. Timescale separation holds by a direct positional measure, though the strict relative comparison is unresolved (Section~\ref{sec:timescale}), and exchange-flux asymmetry holds but is a suggestive analogy, not confirmation of a directed dependency structure (Section~\ref{sec:flux}). The central claim, conditional independence given the blanket, fails across several methods (Section~\ref{sec:conditional_independence}), with measured spatial structure (Section~\ref{sec:corrlength}); neither flux nor boundary concentration gives an effective blanket, and only a band of the surrounding field about one organism radius thick screens (Section~\ref{sec:richer_blanket}).

Thermodynamic and variational free energy share a name but are different mathematical objects, and any connection between them rests on further assumptions. The Friston free energy we computed converges closely with a simple descriptive proxy, so the life-cycle signature is recoverable by more than one route; but both track organism size to a considerable extent (Section~\ref{sec:fe_comparison}), so the agreement is not independent evidence of a deep feature of the chemistry. We could not define an absolute thermodynamic quantity without a further modelling choice (Section~\ref{sec:chemical_fe}); the only comparison made is with the reaction-rate prefactor $uv^2$, a small-$k_r$ leading-order stand-in for entropy production, and its total over the organism is largely a size effect.

\subsection{What Formulation of the Markov Blanket Does This Test?}
\label{sec:blanket_formulation}

Our central test targets the statistical (Pearl-style) structure \citep{Bruineberg2022,Pearl1988}. It bears on the stronger dynamical claim, that interior and exterior influence each other only through the blanket, only to the extent that this claim requires the statistical structure. The test has limits (ESM \S\ref{sec:S_discussion_scope}):
\begin{itemize}
\item $I$ and $E$ are coarse-grained summaries, and the test conditions on flux, not on the thick band that screens.
\item Flux is closer to a collider than to a mediating state (one through which the exterior influences the interior), as the boundary-concentration comparison suggests.
\item We test a single instant, whereas a blanket may more naturally concern history.
\item Our dynamics are deterministic; Friston's construction assumes a stochastic steady-state density.
\item Watershed boundaries are state dependent and could induce dependence; individuating organisms by a fixed threshold on $v$, with no watershed, also finds the violation (ESM \S\ref{sec:S_levelset}) but does not isolate the mechanism.
\end{itemize}
None of these explains the violation away; each limits what it shows.

\subsection{Relation to the Critical Literature}
\label{sec:critical_lit}

Our results sharpen \citeauthor{Raja2021}'s (\citeyear{Raja2021}) concern that a blanket can be imposed on almost any system after the fact, which makes the claims hard to falsify. Used loosely, for any bounded, persisting structure that exchanges material with its environment, ``Markov blanket'' applies to our organisms trivially, and nothing here challenges that usage. Used strictly, to mean that interior and exterior are conditionally independent given the blanket, it makes a claim that can be tested, and we tested it. It fails for the boundary-defined variables we tested: flux and boundary concentration. We tested and rejected one natural rescue: that the violation is an artefact of the step that separates fused spots (it is equally present among organisms that individuate on their own). Friston's own proposal for biologically special blankets, those of minimum entropy with slowly changing constituents \citep{Friston2013}, is met by our organisms by the positional measure of timescale separation, and they still fail to screen with these variables. A band of the surrounding field about one radius thick does screen the interior from the exterior beyond it. We therefore do not claim that no blanket exists, only that a thin, boundary-defined one does not, and the failure is informative because the thick band is cheap: it is what coupling by diffusion produces, and it does not by itself distinguish an organism from any other region of the field. Placebo disks of matched size, placed in the gaps between organisms, are screened too, by a band of 16 px rather than the organisms' 24 px (Section~\ref{sec:richer_blanket}), so the screening belongs to the field; the organisms' extra 8 px of persistence is the one part of the dependence that is theirs.

Is the violation small enough to count as approximate screening \citep{Sakthivadivel2022}? For the boundary-defined variables, no. Flux removes about $38\%$ of the marginal linear dependence, but the nonlinear check shows it \emph{increases} the apparent dependence, and boundary concentration screens only about a fifth of the nonlinear dependence. The conditional coupling has reproducible spatial structure and persists across every individuation method, resolution and estimator tried. For a thick band of the field, yes: below $0.01$ nats. This is compatible with \citeauthor{Aguilera2022}'s (\citeyear{Aguilera2022}) finding that a sparse, thin partition is not generically a blanket, and also with the reply that blankets of finite thickness are to be expected in locally coupled systems. The weight of the inferential reading then rests on the remaining conditions of Section~\ref{sec:fep_intro}: whether synchronisation holds, and whether anything in the dynamics computes about the exterior. The variational free energy remains a valid bound on surprisal whether or not screening is exact. Exact screening matters because it lets that bound be read as encoding beliefs about the exterior (ESM \S\ref{sec:S_discussion_scope}). We do not attribute even approximate inference to this system on the strength of any blanket variable tested.

\subsection{Limitations}
\label{sec:limitations}

Our ergodicity evidence is incomplete, and the ensemble realisations are not fully independent for the population count (Section~\ref{sec:robustness}). The Friston free energy rests on a linear-Gaussian approximation. The chemical free energy was left at the shape level. The results concern the flux, boundary-concentration, near-field-shell and band blanket variables tested and one parameter regime of one model. We did not test whether a parent's state at division predicts its daughter's, so the descriptive continuity across generations is not evidence of heritable internal states. Whether a richer generative model would change the qualitative free-energy results we cannot say, though we would guess not.

\section{Conclusion}
\label{sec:conclusion}

We tested, in a concrete nonlinear pattern-forming system, a central mathematical claim of the free energy principle: that persisting, self-replicating structures possess Markov blankets that render their internal and external states conditionally independent. For Gray--Scott spots, the claim fails for the flux-based blanket we test, under linear and nonlinear estimators, in single-trajectory and ensemble sampling, and under a second individuation method, although timescale separation (by a direct positional measure) and exchange-flux asymmetry (as a suggestive analogy) hold up with qualification. Part of the violation reflects the conditioning variable itself: flux is a joint downstream effect of both sides of the boundary, and conditioning on it raises the dependence, as expected of a collider. Conditioning on the boundary's own concentration does only partial work. No boundary-defined variable tested here, flux included, gives an effective blanket. A band of the surrounding field about one organism radius thick does screen the interior from the exterior beyond it, to under $0.01$ nats; that is what short-range coupling produces in any local field, not evidence of inference. A variational free energy from an explicit generative model can be computed for these organisms and gives a life-cycle signature that agrees closely with a simple descriptive proxy, partly through a shared dependence on organism size; this reflects well-behaved chemical dynamics, not Bayesian inference. A chemically motivated memory extension, which might have been expected to support better individuation, does not remove the violation. Self-replicating reaction-diffusion spots are simple enough to simulate exhaustively and to individuate operationally, which makes them a direct testbed for questions about the scope of the free energy principle that are otherwise addressed through toy linear systems or untested assertions about real biology.

\section*{Ethics}

Not applicable.

\section*{Data Accessibility}

Code for the solvers, the individuation and lineage-tracking pipeline, the conditional mutual information estimator, and the scripts for every figure and table is available at \url{https://github.com/johnepearson/GS_FEP}, together with the settled starting field from which every simulation is run and a README giving the command for each result.

\section*{Authors' Contributions}

J.E.P. conceived the study, designed and interpreted all analyses, and wrote the paper.

\section*{Competing Interests}

The author declares no competing interests.

\section*{Funding}

This work received no external funding.

\section*{Acknowledgements}

Claude (Anthropic) was used extensively for developing and debugging code, running the numerical analyses, and drafting and revising text. All results were reviewed, verified, and are the responsibility of the author. AI-generated reviews from Anthropic, OpenAI, and Google Gemini were employed repeatedly. We are especially grateful to Mac Hyman, who submitted two drafts of the manuscript to a group of five OpenAI reviewers he developed. Andy Fraser pointed out that the coarse-graining into $I$ and $E$ is justified by the data processing inequality (ESM \S\ref{sec:S_coarse}).

\bibliographystyle{plainnat}
\bibliography{references}

\begin{thebibliography}{26}
\providecommand{\natexlab}[1]{#1}
\providecommand{\url}[1]{\texttt{#1}}
\expandafter\ifx\csname urlstyle\endcsname\relax
  \providecommand{\doi}[1]{doi: #1}\else
  \providecommand{\doi}{doi: \begingroup \urlstyle{rm}\Url}\fi

\bibitem[Aguilera et~al.(2022)Aguilera, Millidge, Tschantz, and
  Buckley]{Aguilera2022}
Miguel Aguilera, Beren Millidge, Alexander Tschantz, and Christopher~L.
  Buckley.
\newblock How particular is the physics of the free energy principle?
\newblock \emph{Physics of Life Reviews}, 40:\penalty0 24--50, 2022.
\newblock \doi{10.1016/j.plrev.2021.11.001}.

\bibitem[Beer(2015)]{Beer2015}
Randall~D. Beer.
\newblock Characterizing autopoiesis in the {G}ame of {L}ife.
\newblock \emph{Artificial Life}, 21\penalty0 (1):\penalty0 1--19, 2015.
\newblock \doi{10.1162/ARTL_a_00143}.

\bibitem[Beer(2020)]{Beer2020}
Randall~D. Beer.
\newblock An integrated perspective on the constitutive and interactive
  dimensions of autonomy.
\newblock In \emph{Proceedings of the 2020 Conference on Artificial Life (ALIFE
  2020)}, pages 202--209. MIT Press, 2020.

\bibitem[Biehl et~al.(2016)Biehl, Ikegami, and Polani]{Biehl2016}
Martin Biehl, Takashi Ikegami, and Daniel Polani.
\newblock Towards information based spatiotemporal patterns as a foundation for
  agent representation in dynamical systems.
\newblock In \emph{Proceedings of the Artificial Life Conference 2016}. MIT
  Press, 2016.

\bibitem[Bruineberg et~al.(2022)Bruineberg, Do{\l}{\k{e}}ga, Dewhurst, and
  Baltieri]{Bruineberg2022}
Jelle Bruineberg, Krzysztof Do{\l}{\k{e}}ga, Joe Dewhurst, and Manuel Baltieri.
\newblock The emperor's new {M}arkov blankets.
\newblock \emph{Behavioral and Brain Sciences}, 45:\penalty0 e183, 2022.
\newblock \doi{10.1017/S0140525X21002351}.

\bibitem[Cover and Thomas(2006)]{CoverThomas2006}
Thomas~M. Cover and Joy~A. Thomas.
\newblock \emph{Elements of Information Theory}.
\newblock Wiley-Interscience, Hoboken, NJ, 2nd edition, 2006.

\bibitem[Da~Costa et~al.(2021)Da~Costa, Friston, Heins, and
  Pavliotis]{DaCosta2021}
Lancelot Da~Costa, Karl Friston, Conor Heins, and Grigorios~A. Pavliotis.
\newblock {B}ayesian mechanics for stationary processes.
\newblock \emph{Proceedings of the Royal Society A}, 477:\penalty0 20210518,
  2021.
\newblock \doi{10.1098/rspa.2021.0518}.

\bibitem[Frenzel and Pompe(2007)]{FrenzelPompe2007}
Stefan Frenzel and Bernd Pompe.
\newblock Partial mutual information for coupling analysis of multivariate time
  series.
\newblock \emph{Physical Review Letters}, 99\penalty0 (20):\penalty0 204101,
  2007.
\newblock \doi{10.1103/PhysRevLett.99.204101}.

\bibitem[Friston(2010)]{Friston2010}
Karl Friston.
\newblock The free-energy principle: A unified brain theory?
\newblock \emph{Nature Reviews Neuroscience}, 11\penalty0 (2):\penalty0
  127--138, 2010.
\newblock \doi{10.1038/nrn2787}.

\bibitem[Friston(2013)]{Friston2013}
Karl Friston.
\newblock Life as we know it.
\newblock \emph{Journal of the Royal Society Interface}, 10\penalty0
  (86):\penalty0 20130475, 2013.
\newblock \doi{10.1098/rsif.2013.0475}.

\bibitem[Friston(2019)]{Friston2019Particular}
Karl Friston.
\newblock A free energy principle for a particular physics.
\newblock \emph{arXiv preprint arXiv:1906.10184}, 2019.

\bibitem[Friston et~al.(2006)Friston, Kilner, and Harrison]{Friston2006}
Karl Friston, James Kilner, and Lee Harrison.
\newblock A free energy principle for the brain.
\newblock \emph{Journal of Physiology-Paris}, 100\penalty0 (1-3):\penalty0
  70--87, 2006.
\newblock \doi{10.1016/j.jphysparis.2006.10.001}.

\bibitem[Friston et~al.(2021)Friston, Heins, Ueltzh{\"o}ffer, Da~Costa, and
  Parr]{Friston2021Stochastic}
Karl Friston, Conor Heins, Kai Ueltzh{\"o}ffer, Lancelot Da~Costa, and Thomas
  Parr.
\newblock Stochastic chaos and {M}arkov blankets.
\newblock \emph{Entropy}, 23\penalty0 (9):\penalty0 1220, 2021.
\newblock \doi{10.3390/e23091220}.

\bibitem[Friston et~al.(2023)Friston, Da~Costa, Sajid, Heins, Ueltzh{\"o}ffer,
  Pavliotis, and Parr]{Friston2023Simpler}
Karl Friston, Lancelot Da~Costa, Noor Sajid, Conor Heins, Kai Ueltzh{\"o}ffer,
  Grigorios~A. Pavliotis, and Thomas Parr.
\newblock The free energy principle made simpler but not too simple.
\newblock \emph{Physics Reports}, 1024:\penalty0 1--29, 2023.
\newblock \doi{10.1016/j.physrep.2023.07.001}.

\bibitem[Gray and Scott(1983)]{Gray1983}
P.~Gray and S.~K. Scott.
\newblock Autocatalytic reactions in the isothermal, continuous stirred tank
  reactor: Isolas and other forms of multistability.
\newblock \emph{Chemical Engineering Science}, 38\penalty0 (1):\penalty0
  29--43, 1983.
\newblock \doi{10.1016/0009-2509(83)80132-8}.

\bibitem[Gray and Scott(1984)]{GrayScott1984}
P.~Gray and S.~K. Scott.
\newblock Autocatalytic reactions in the isothermal, continuous stirred tank
  reactor: Oscillations and instabilities in the system a + 2b $\rightarrow$
  3b, b $\rightarrow$ c.
\newblock \emph{Chemical Engineering Science}, 39\penalty0 (6):\penalty0
  1087--1097, 1984.
\newblock \doi{10.1016/0009-2509(84)87017-7}.

\bibitem[Kraskov et~al.(2004)Kraskov, St\"ogbauer, and
  Grassberger]{Kraskov2004}
Alexander Kraskov, Harald St\"ogbauer, and Peter Grassberger.
\newblock Estimating mutual information.
\newblock \emph{Physical Review E}, 69\penalty0 (6):\penalty0 066138, 2004.
\newblock \doi{10.1103/PhysRevE.69.066138}.

\bibitem[Mahara et~al.(2004)Mahara, Suematsu, Yamaguchi, Ohgane, Nishiura, and
  Shimomura]{Mahara2004}
Hitoshi Mahara, Nobuhiko~J. Suematsu, Tomohiko Yamaguchi, Kunishige Ohgane,
  Yasumasa Nishiura, and Masatsugu Shimomura.
\newblock Three-variable reversible {G}ray-{S}cott model.
\newblock \emph{Journal of Chemical Physics}, 121\penalty0 (18):\penalty0
  8968--8972, 2004.
\newblock \doi{10.1063/1.1803531}.

\bibitem[Munafo(2022)]{Munafo}
Robert~P. Munafo.
\newblock {P}earson's classification (extended) of {G}ray-{S}cott system
  parameter values at {MROB}.
\newblock \url{http://www.mrob.com/pub/comp/xmorphia/pearson-classes.html},
  March 2022.
\newblock Accessed 10 October 2026.

\bibitem[Otsu(1979)]{Otsu1979}
Nobuyuki Otsu.
\newblock A threshold selection method from gray-level histograms.
\newblock \emph{IEEE Transactions on Systems, Man, and Cybernetics}, 9\penalty0
  (1):\penalty0 62--66, 1979.
\newblock \doi{10.1109/TSMC.1979.4310076}.

\bibitem[Palacios et~al.(2020)Palacios, Razi, Parr, Kirchhoff, and
  Friston]{Palacios2020}
Ensor~Rafael Palacios, Adeel Razi, Thomas Parr, Michael Kirchhoff, and Karl
  Friston.
\newblock On {M}arkov blankets and hierarchical self-organisation.
\newblock \emph{Journal of Theoretical Biology}, 486:\penalty0 110089, 2020.
\newblock \doi{10.1016/j.jtbi.2019.110089}.

\bibitem[Pearl(1988)]{Pearl1988}
Judea Pearl.
\newblock \emph{Probabilistic Reasoning in Intelligent Systems: Networks of
  Plausible Inference}.
\newblock Morgan Kaufmann, San Mateo, CA, 1988.

\bibitem[Pearson(1993)]{Pearson1993}
John~E. Pearson.
\newblock Complex patterns in a simple system.
\newblock \emph{Science}, 261\penalty0 (5118):\penalty0 189--192, 1993.
\newblock \doi{10.1126/science.261.5118.189}.

\bibitem[Raja et~al.(2021)Raja, Valluri, Baggs, Chemero, and
  Anderson]{Raja2021}
Vicente Raja, Dinesh Valluri, Edward Baggs, Anthony Chemero, and Michael~L.
  Anderson.
\newblock The {M}arkov blanket trick: On the scope of the free energy principle
  and active inference.
\newblock \emph{Physics of Life Reviews}, 39:\penalty0 49--72, 2021.
\newblock \doi{10.1016/j.plrev.2021.09.001}.

\bibitem[Runge(2018)]{Runge2018}
Jakob Runge.
\newblock Conditional independence testing based on a nearest-neighbor
  estimator of conditional mutual information.
\newblock In \emph{Proceedings of the Twenty-First International Conference on
  Artificial Intelligence and Statistics}, volume~84, pages 938--947. PMLR,
  2018.

\bibitem[Sakthivadivel(2022)]{Sakthivadivel2022}
Dalton A.~R. Sakthivadivel.
\newblock Weak {M}arkov blankets in high-dimensional, sparsely-coupled random
  dynamical systems, 2022.
\newblock arXiv:2207.07620.

\end{thebibliography}

\end{document}